\documentclass[sigconf]{acmart}

\AtBeginDocument{%
  \providecommand\BibTeX{{%
    \normalfont B\kern-0.5em{\scshape i\kern-0.25em b}\kern-0.8em\TeX}}}

\setcopyright{none}
\copyrightyear{2020}
\acmYear{2020}
\acmDOI{.}

\acmJournal{TOCT}

\usepackage[all]{xy}
\usepackage{flushend}
\usepackage{balance}
\newcommand{\blue}[1]{\textcolor[rgb]{0,0,1}{#1}}
\newcommand{\vect}[1]{\boldsymbol{#1}}

\begin{document}

\title{Information Flow in Color Appearance Neural Networks}

\author{Jes{\'u}s Malo}
\email{jesus.malo@uv.es}
\email{http://isp.uv.es/}
\orcid{0000-0002-5684-8591}
\affiliation{%
  \institution{Image Processing Lab, Universitat de Valencia}
  \streetaddress{9 Catedratico Escadino}
  \city{46980 Paterna}
  \state{Valencia}
  \country{Spain}
}

\renewcommand{\shortauthors}{J. Malo}

\begin{abstract}
Color Appearance Models are biological networks that consist of a cascade of linear+nonlinear layers that
modify the linear measurements at the retinal photo-receptors leading to an
internal (nonlinear) representation of color that correlates with psychophysical experience.
The basic layers of these networks include:
(1)~chromatic adaptation (normalization of the mean and covariance of the color manifold),
(2)~change to opponent color channels (PCA-like rotation in the color space), and
(3)~saturating nonlinearities to get perceptually Euclidean color representations (similar to dimension-wise equalization).
The \emph{Efficient Coding Hypothesis} argues that these transforms should emerge from information-theoretic goals.
In case this hypothesis holds in color vision, the question is, \emph{what is the coding gain due to the different layers of the color appearance networks?}

In this work, a representative family of Color Appearance Models is analyzed in terms of how the redundancy among the
chromatic components is modified along the network and how much information is transferred from the input data to the noisy response.
The proposed analysis is done using data and methods that were not available before:
(1) new colorimetrically calibrated scenes in different CIE illuminations for
proper evaluation of chromatic adaptation, and
(2) new statistical tools to estimate (multivariate) information-theoretic quantities between multidimensional sets
based on Gaussianization.
Results confirm that the Efficient Coding Hypothesis holds for current color vision models,
and identify the psychophysical mechanisms critically responsible for gains in information
transference: opponent channels and their nonlinear nature are more important than chromatic
adaptation at the retina.

\end{abstract}



\maketitle

\section{Introduction}
Biological vision is relevant to manage visual data because
natural visual systems evolved to develop efficient representations of visual features
that may be an inspiration for artificial systems. Examples include (1)~the
equivalence between the spatio-spectral sensitivity of receptive fields in natural
neural systems and those emerging from information maximization and optimal matching \cite{Gutmann14},
or the filters found by maximizing classification performance \cite{AlexNet12}, and
(2)~the importance of human-like spatio-chromatic representations in image
coding algorithms \cite{Marcellin01,gutierrez2012color}.

Conversely, the quantitative tools of statistical learning are key to propose principled theories in visual neuroscience.
For instance the classical \emph{Efficient Coding Hypothesis} argues that the organization of
biological sensors comes from the optimization of information-theoretic goals \cite{Barlow59,Barlow01}.
The conventional approach to check this hypothesis is \emph{from-statistics-to-perception}: i.e. deriving the biological behavior
from statistical arguments. In color vision, this includes
the derivation of opponent channels from PCA \cite{Buchsbaum83},
and the derivation of the nonlinearities of opponent channels \cite{MacLeod03b,Laparra12},
and even the reproduction of color illusions \cite{Laparra15}, from information maximization or error minimization
arguments.
However, there is an alternative way to check the hypothesis: \emph{from-perception-to-statistics} \cite{Malo10}.
In this case, perceptually meaningful models which have not been statistically optimized are shown to have a statistically efficient behavior.

In this work we take this alternative approach (\emph{from-perception-to-statistics}) for color vision models.
Here we analyze the communication efficiency of standard color vision models (that already describe a wide range
of color vision psychophysics) using multivariate information-theoretic measurements.
This analysis is interesting for data science and computer vision because the building blocks of color appearance models
have received statistical interpretation: adaptation is related to manifold alignment \cite{Webster97},
opponency is related to principal components of color data \cite{Buchsbaum83}, and
the nonlinearities of the opponent channels are related to histogram equalization \cite{MacLeod03b,Laparra12,Laparra15}.
In case the efficient coding hypothesis holds for color vision networks, the question is,
\emph{what is the coding gain due to the different layers of the color appearance networks?}

A quantitative response to that question is interesting for data science to select which process should be
addressed first for efficient management of color data.
The \emph{perception-to-statistics} approach has been previously applied to texture perception \cite{Malo10,Gomez19},
but it is original in color vision: note that previous information-theoretic
analysis of color vision (e.g. \cite{FosterJOSA18} and references therein) were focused on the
amount of information from an image which can be obtained from the corresponding scene under a different illumination,
and not on the efficiency of the system to transmit generic color information (irrespective of the illumination)
as done here.

\section{Statistical interpretation of building blocks of color vision}

The basic elements of biological color vision are:

\textbf{(1) Linear integration of spectral irradiance at the retina} by three sensors tuned to \emph{long}, \emph{medium} and \emph{short} wavelengths, which are commonly referred to as LMS sensors \cite{Stockman11}. In every natural or artificial system this initial linear stage is the necessary transduction from electromagnetic energy to the first numerical representation of color data.
Here we will start by expressing colorimetrically calibrated images in LMS tristimulus values via the Stockman and Sharpe fundamentals \cite{Stockman00}.

\textbf{(2) Nonlinear adaptation at the retina} adjusts the sensitivity (or gain) of the LMS sensors to the illumination of the scene.
For instance, classical Von-Kries chromatic adaptation normalizes the sensors by the responses of what is considered to
be \emph{white} in the scene \cite{Fairchild13}. From a statistical point of view, the role of chromatic adaptation is the same as manifold alignment in machine learning to make interpretation of the data easier in changing environments (in this case, environments with different illuminations).
According to this interpretation, generalizations of the Von-Kries transform have been proposed, as for instance trying to make first and second moments of the different color manifolds equal \cite{Webster97,Clifford07}, or using higher order equalization transforms for the different datasets and making their dimensions equal in the canonical domain. The latter higher order methods may be linear \cite{Gutmann14} or nonlinear \cite{Laparra12,Laparra15}.
Here we will explore the behavior of classical Von-Kries adaptation \cite{Fairchild13}, and the adaptation through the equalization of mean and covariance, which will be referred to as the Webster-Clifford approach following \cite{Webster97,Clifford07}. More sophisticated nonlinear equalization techniques such as the Sequential Principal Curves Analysis (SPCA) \cite{Laparra12,Laparra15} will be used as a convenient statistical benchmark.

\textbf{(3) Linear opponent channels in ganglion cells and beyond}.
The change from a color representation mediated by sensors with \emph{all-positive} (physically realizable) sensitivities, as the LMS sensors at the retina, to color representations in \emph{opponent channels} is obtained via linear recombination of the LMS signals.
This recombination leads to an achromatic channel and two red-green and yellow-blue chromatic channels \cite{Jameson59,Stockman11,Fairchild13}.
In this way, neural computation allows to obtain sensors with \emph{opponent} (positive-and-negative) spectral sensitivities that are not easy to implement physically. Spectral sensitivities which are effectively opponent are found at different layers along the neural pathway: at the ganglion cells, the lateral geniculate nucleus, and the visual cortex \cite{Shapley11}.
This linear change of color representation has been statistically interpreted as the identification of principal components
of the color manifold \cite{Buchsbaum83}.
Here we will use the classical Jameson and Hurvich transform to opponent channels \cite{Jameson59}, which is not based on image statistics, but on color matching experiments. These opponent channels are also called \emph{achromatic}, \emph{tritanopic} and \emph{deuteranopic} (ATD) due to their relation with dichromatic vision \cite{Stockman11,Capilla04}.

\textbf{(4) Nonlinearities of opponent channels}.
Nonuniform color discrimination thresholds \cite{Krauskopf92,Romero93}
imply that the saturation of the Weber law occurring in the achromatic channel \cite{Stiles82,Fairchild13} also appears in the chromatic channels \cite{Stockman11}. This nonlinearity of the opponent channels has been explained as the necessary transform to get equalized PDFs in the responses \cite{MacLeod03b}. This nonlinearity can be optimized for PDF equalization or error minimization after uniform quantization. The authors referred to this nonlinearity as the Pleistochrome transform \cite{MacLeod03b}.
This dimension-wise equalization concept was generalized to multivariate scenarios through principal curves (the above mentioned SPCA) \cite{Laparra12,Laparra15}.
Here, the simpler (univariate) pleistochrome transform will be compared to the more general SPCA transform.

The sequence of the building blocks considered above fits into the current deep-network paradigm \cite{Goodfellow16} because it can be implemented as a cascade of two \emph{linear+nonlinear} layers performing \emph{spectral integration+adaptation} followed by \emph{opponency+saturation}:
\begin{equation}
  \xymatrixcolsep{2pc}
  \xymatrix{ \vect{x}^0 \ar@/^1pc/[r]^{\scalebox{1.00}{$\mathcal{L}^{(1)}$}} & \vect{r}^1  \ar@/^1pc/[r]^{\scalebox{1.00}{$\mathcal{N}^{(1)}$}} & \vect{x}^1 \ar@/^1pc/[r]^{\scalebox{1.00}{$\mathcal{L}^{(2)}$}}  & \vect{r}^2 \ar@/^1pc/[r]^{\scalebox{1.00}{$\mathcal{N}^{(2)}$}} & \vect{x}^2
  }
  \label{modular}
\end{equation}
In this architecture, $\vect{x}^0$ is the spectral irradiance at the photo-receptors,
$\vect{r}^1$ is the vector of linear LMS responses,
$\vect{x}^1$ is the vector of nonlinearly adapted LMS responses,
$\vect{r}^2$ is the vector of linearly recombined ATD responses,
and $\vect{x}^2$ is the vector of nonlinear ATD responses.
The goal of the work is measuring how much information from the linear LMS stage is transmitted to
the other layers of the network. In this way we check if information transmission can be a
sensible organization principle for these natural systems and which are the critical layers of the
network.

\section{Materials and Methods}

\subsection{Colorimetrically calibrated database}
\begin{figure}[h]
  \centering
  \includegraphics[width=\linewidth]{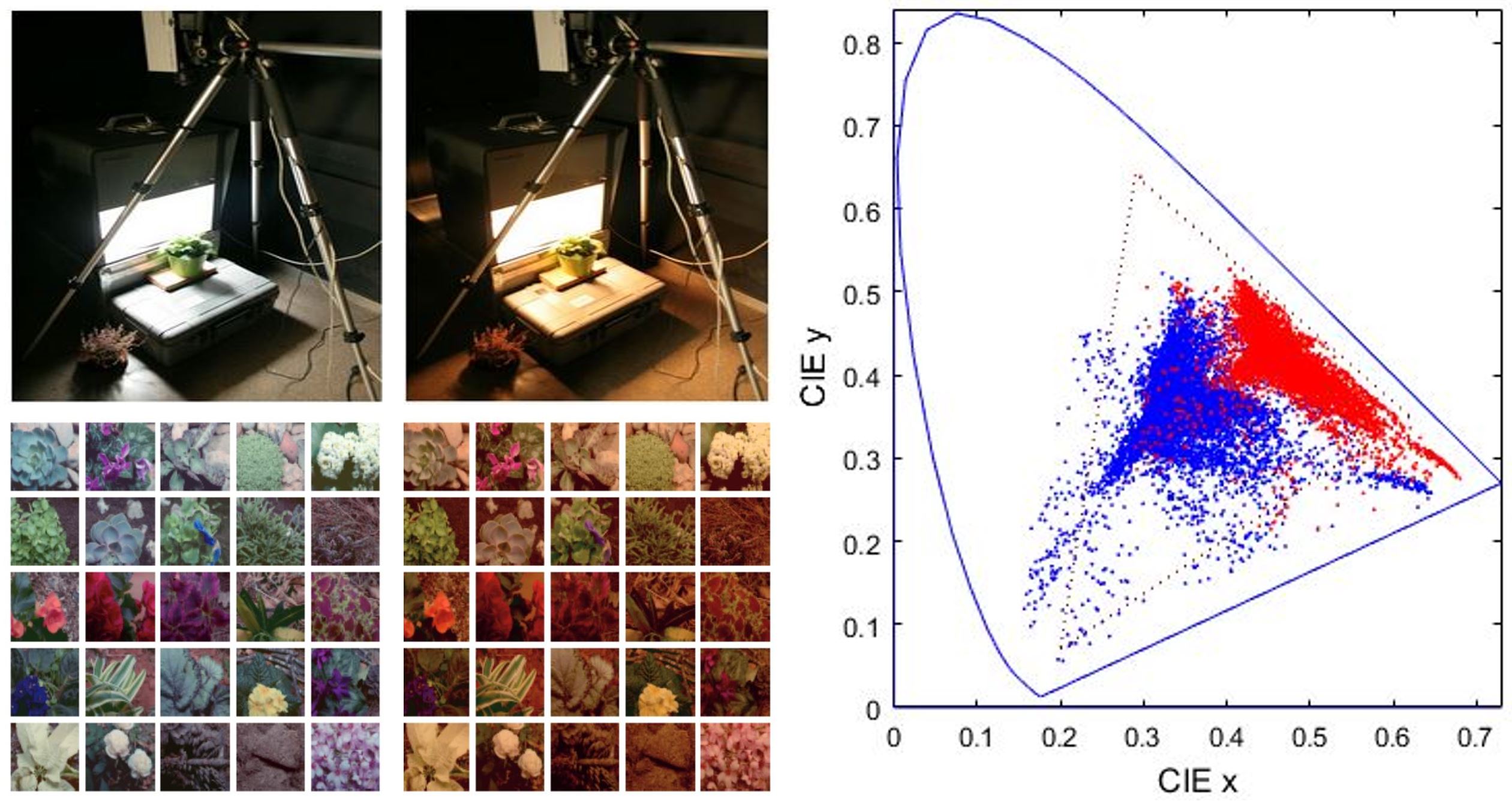}
  \caption{\small{Natural colors used in the experiments.
  \emph{Left:} experimental setting to record the natural scenes under calibrated illumination (CIE D65 and CIE A spectra) and representative corresponding scenes.
  \emph{Right:} color measurements in the CIE xy diagram. Blue and red dots represent the colors under D65 (white) and A (yellowish) illuminations.}}
  \Description{Data used in the experiments.}
  \label{data}
\end{figure}

The IPL color image database \cite{Laparra12,Gutmann14,Laparra15} is well suited to study color adaptation because its controlled illumination under CIE A and CIE D65 allows us to know the white point in each scene. This implies that chromatic adaptation transforms will not require extra approximations such as the gray-world assumption. Acquisition of the controlled images and resulting data is illustrated in Fig.~\ref{data}.

\subsection{Total Correlation and Mutual Information from Gaussianization}
Total Correlation, $T$, \cite{Watanabe60,Studeny98}
is required to quantify the redundancy within the layers of the considered networks; and Mutual Information, $I$, \cite{Cover06}, is required to quantify the amount of transferred information from one layer of the network to the next.
A possible measure of the success in the chromatic adaptation is the Kullback-Leibler divergence, \emph{KLD}, \cite{Cover06} between
the color distributions after illumination compensation.
However, estimation of these (multivariate) quantities is not straightforward: naive use of their definition implies the estimation of multivariate PDFs and this would introduce substantial bias in the results.

Here we solve the above problem by using a novel estimator of $T$ which only relies on (easier) univariate density estimations: the Rotation-Based Iterative Gaussianization (RBIG) \cite{ICLR19}.
The RBIG is a cascade of nonlinear+linear layers, each one made of (easy) marginal Gaussianizations followed by an (easy) rotation. This invertible architecture is able to transform any input PDF into a zero-mean unit-covariance multivariate Gaussian even if the chosen rotations are random \cite{Laparra11}.
This ability to completely remove the structure of any PDF is useful to estimate $T$ of arbitrary vectors $\vect{x}$: as the redundancy of a Gaussianized signal is zero, $T(\vect{x})$ corresponds to the sum of the individual variations that take place along the layers of RBIG while Gaussianizing $\vect{x}$. Interestingly, the individual variation in each RBIG layer only depends on (easy to compute) univariate negentropies \cite{Laparra11}.
The information shared by multidimensional datasets, $I$, is just the remaining total correlation within the variables once they have been separately Gaussianized  \cite{ICLR19}. Similarly, the \emph{KLD} between two multidimensional variables is just the marginal negentropy of one of the variables when it has been Gaussianized using the Gaussianization transform learnt from the other. Therefore, both $I$ and $KLD$ can also be estimated using Gaussianization.

We also computed $T$ and $I$ via the Kozachenko-Leonenko estimator \cite{Kozachenko87} used in other studies on color data \cite{Ivan13}.
However, note that this alternative estimator cannot be used to compute the \emph{KLD}.

\section{Experiments and Results}

In this work we study the communication efficiency of three representative families of color vision models:
\emph{\textbf{(a)}}~cascades of physiologically meaningful linear+nonlinear layers: (i)~LMS sensors \cite{Stockman00} + Von-Kries or Webster-Clifford adaptation \cite{Webster97}, followed by (ii)~Jameson \& Hurvich linear opponent channels \cite{Jameson59} + plesitochrome saturations \cite{MacLeod03b}.
\emph{\textbf{(b)}}~standard color appearance models that are made of the same ingredients, including (i) the classical CIE Lab model \cite{CIELab76}, (ii) the L Lab model \cite{LLab96}, and (iii) the more recent CIECAM02 \cite{ciecam97,ciecam02}; and finally,
\emph{\textbf{(c)}}~as a convenient statistically-based benchmark, we will also measure the performance of our multivariate equalization technique  based on principal curves, SPCA \cite{Laparra12,Laparra15}, with or without classical chromatic adaptation transforms.

\begin{figure*}[h]
  \centering
  \includegraphics[width=\linewidth]{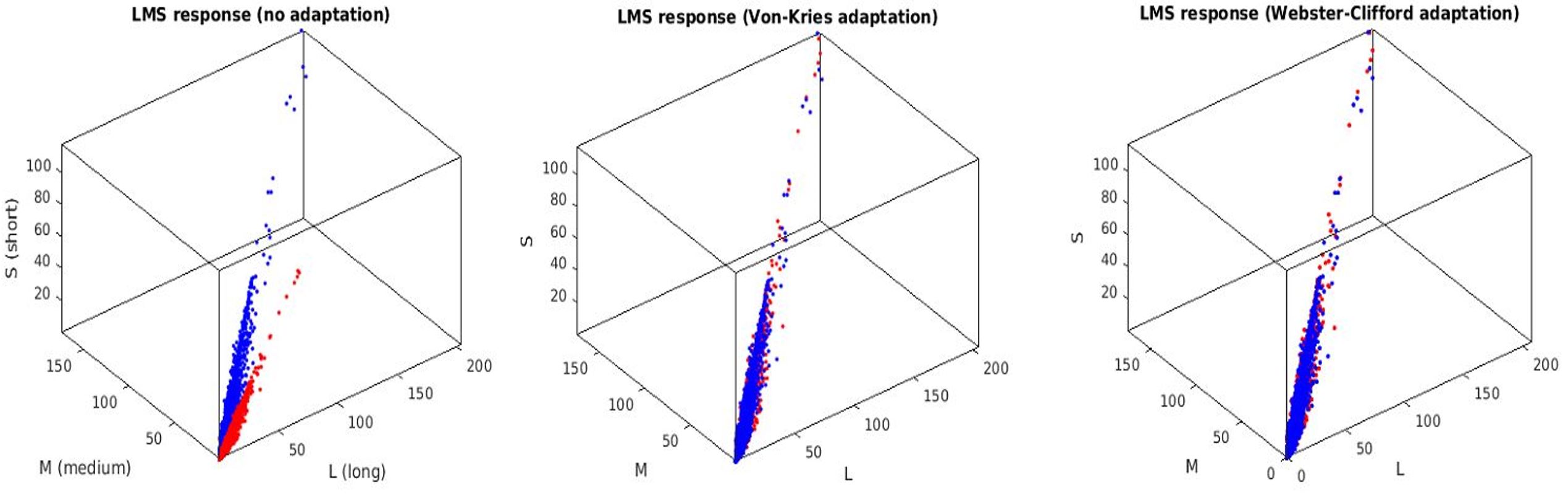}\\
  \includegraphics[width=\linewidth]{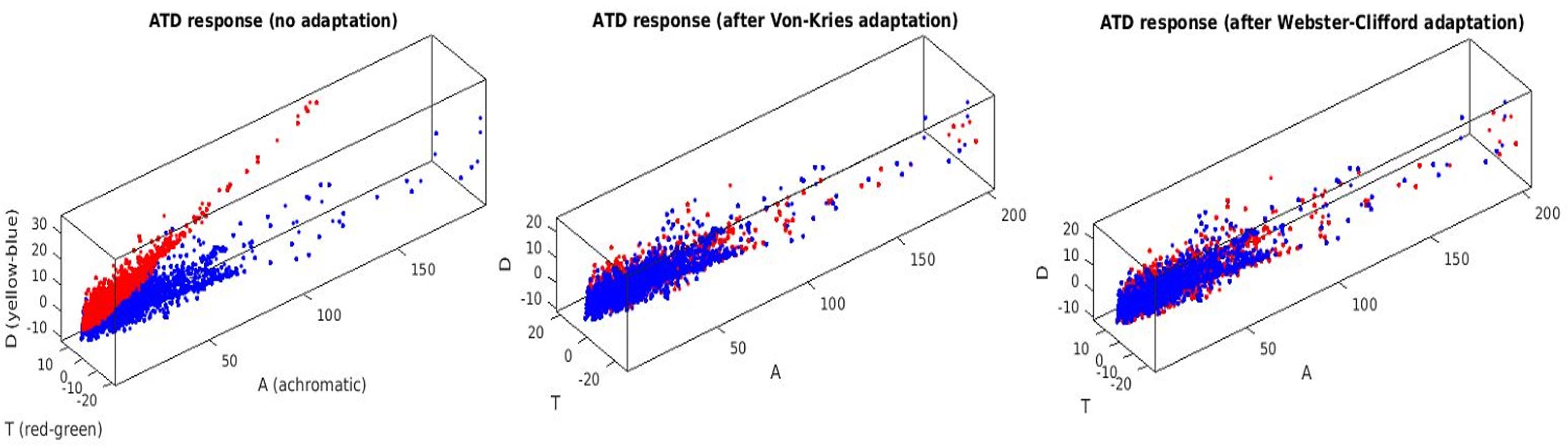}\\
  \includegraphics[width=1.01\linewidth]{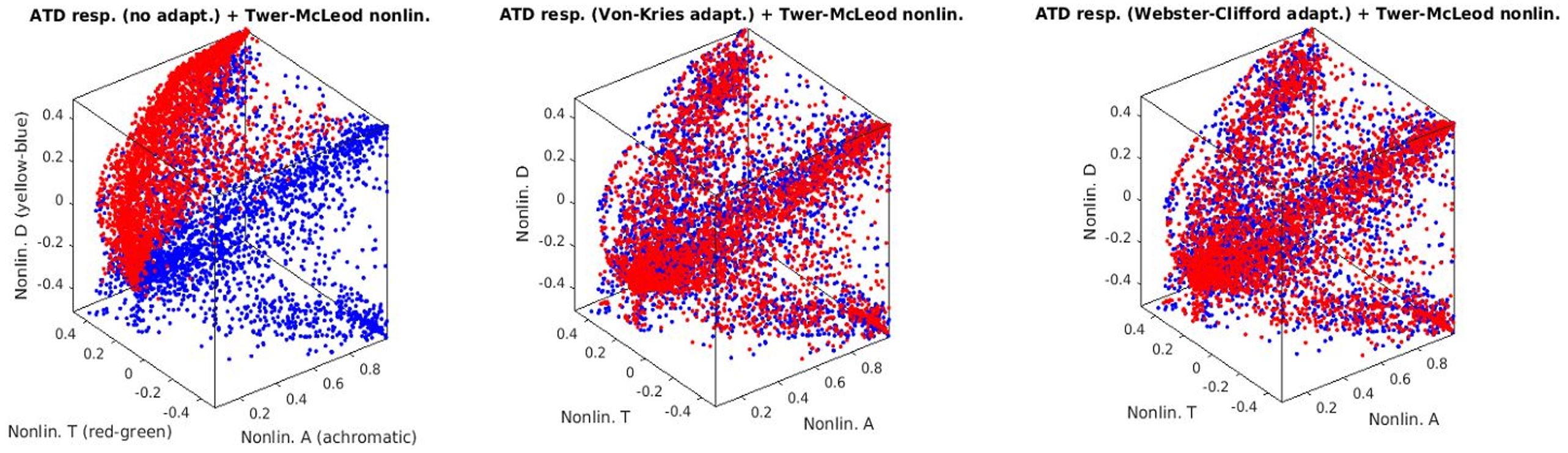}
  \caption{\small{Manifold changes through physiological networks. \emph{Top Row:} Retinal responses of LMS cones with no adaptation (left) and with different retinal adaptation schemes (center and right). \emph{Center Row:} Linear opponent ATD responses (recombination after the retina) with no adaptation (left) and different adaptation schemes. \emph{Bottom Row:} Nonlinear opponent ATD responses after dimension-wise PDF equalization (pleistochrome transform). Red and blue dots represent the samples under the A and D65 illuminations.}}
  \Description{Physiological representations.}
  \label{fig_physiol}
\end{figure*}

\begin{figure*}[h]
  \centering
  \includegraphics[width=\linewidth]{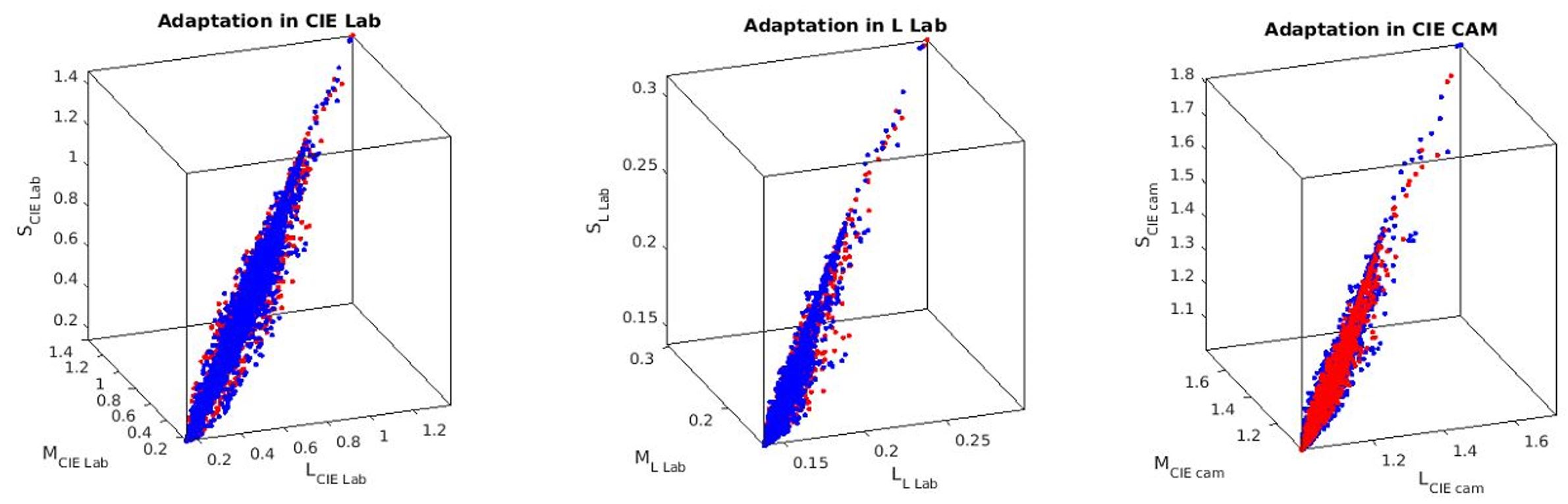}\\
  \includegraphics[width=\linewidth]{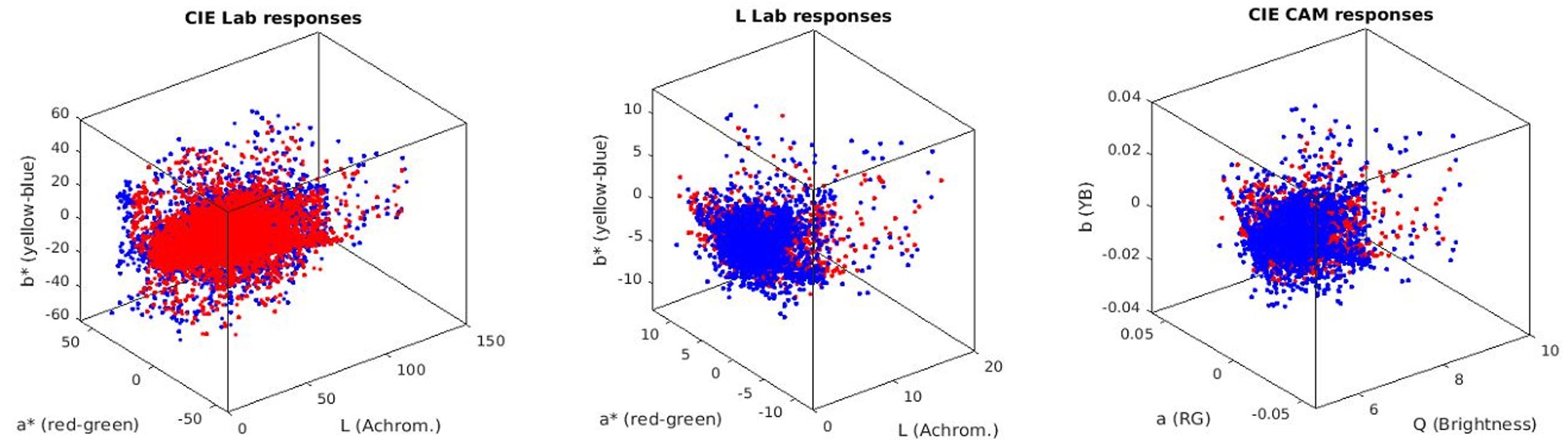}
  \caption{\small{Manifold changes through psychophysical color appearance models. \emph{Top:} Responses of sensors of standard models
  tuned to Long, Medium and Short wavelengths. All these representations already include divisive adaptation schemes for manifold alignment. \emph{Bottom:} Nonlinear opponent responses in the different color appearance models. Red and blue dots represent the samples under the A and D65 illuminations.}}
  \Description{CAMS.}
  \label{fig_cams}
\end{figure*}

\begin{figure*}[h]
  \centering
  \includegraphics[width=\linewidth]{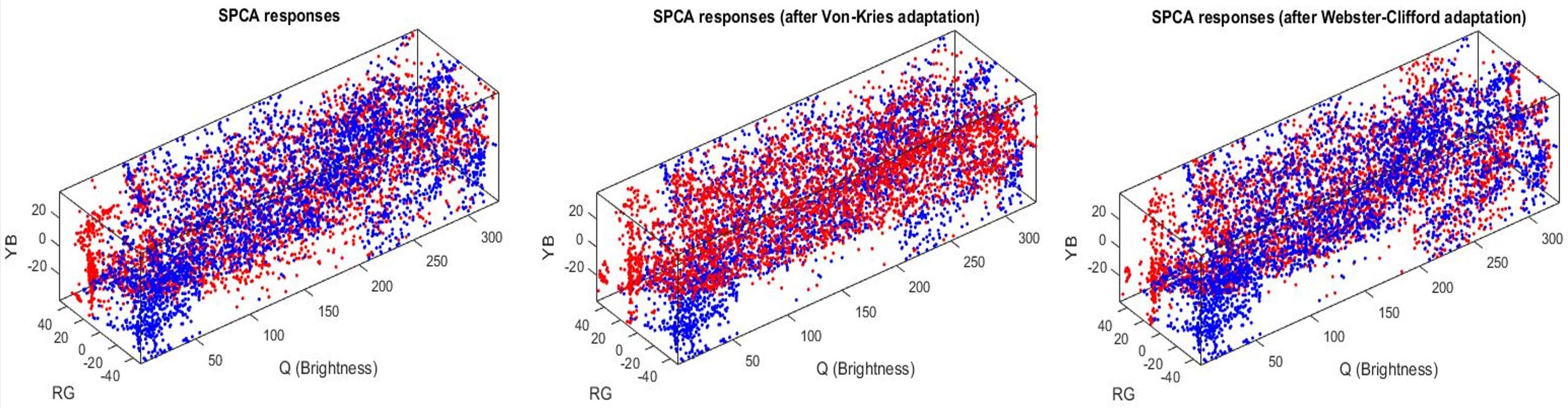}
  \caption{\small{Manifold equalization using nonlinear ICA (Sequential Principal Curves, SPCA). \emph{Left:} Responses of the SPCA sensors from non-aligned manifolds. \emph{Center:} SPCA responses from Von-Kries adapted measurements. \emph{Right:} SPCA responses from Webster-Clifford adapted colors. Red and blue dots represent the samples under the A and D65 illuminations.}}
  \Description{SPCAS.}
  \label{fig_spcas}
\end{figure*}

\paragraph{\textbf{Visualization of the color manifolds.}}
First, we show how the color manifolds of natural scenes under the CIE D65 and CIE A illuminations
change through the layers of the considered networks.
Information measures are related to the volume and density of the data distribution \cite{Cover06}.
As a result, visualization of the shape of the manifolds is important to understand the geometric effect of the considered transforms,
and hence, their impact in the measures.
Specifically, Fig. \ref{fig_physiol} shows the separate effect of the elements of the physiological networks (opponency and nonlinearities of opponent channels without and with different chromatic adaptations).
Fig.~\ref{fig_cams} shows the effect of different (psychophysical) Color Appearance Models once the corresponding chromatic adaptation has
been performed and after the nonlinear opponent responses are obtained.
Finally, Fig.~\ref{fig_spcas} shows the result of the nonlinear equalization SPCA when it is applied to nonaligned manifolds or to manifolds aligned according to different color adaptation transforms.

\paragraph{\textbf{Quantification of the information flow.}}
We provide indirect measures of efficiency, such as the redundancy, $T$, within each layer along the networks, and the differential entropy, $H$, of the signal at each layer of the networks (Tables \ref{table_T} and \ref{table_H} respectively).
We also provide a direct measure of efficiency: the mutual information between the noiseless input and the noisy response, $I$ at each layer of the network (Table \ref{table_I}).

These magnitudes were estimated from the responses of the models to test sets of randomly chosen color samples from the IPL database shown in Fig. \ref{data}.
The parameters of the models that require statistical training (namely the Webster-Clifford adaptation, the pleistochrome nonlinearity, and SPCA)
were trained over a set with $2 \cdot 10^6$ natural color samples. Then, $4 \cdot 10^4$ color samples not included in the training set were transformed using the perceptual models and the purely-statistical SPCA.
10 independent estimations of the quantities were done using random subsets of 80\% of the test set.
All tables shown were estimated using RBIG. Additionally, entropy and mutual information were also estimated using the Kozachenko-Leonenko procedure
leading to qualitatively similar results (tables equivalent to Table \ref{table_H} and \ref{table_I}, not shown).
Given the dependence of $H$ with arbitrary scaling of the PDF support, for fair comparison, computation of $H$ in Table \ref{table_H} was done after
linearly re-scaling the signals to be inscribed in the same cube of size 10.
In this way, the differential entropy values do describe how uniform the distributions are in the common support.
As a useful reference, note that the theoretical differential entropy of the uniform distribution in this domain (the upper bound for $H$) is 9.97 bits.

\begin{table}
  \caption{Intra-layer Total Correlation through the networks (in bits). Optimal systems should have $T = 0$ bits.}
  \vspace{-0.3cm}
  \label{table_T}
  \begin{tabular}{lccc}
      \toprule
    \small{\textbf{Physiol. Models}} & \small{No adaptation} & \small{Von-Kries} & \small{Webster-Clifford} \\
    \midrule
    \small{Linear LMS input}    & $5.71 \pm 0.04$ & $6.26 \pm 0.02$ & $6.29 \pm 0.03$ \\
    \small{Linear ATD channels} & $3.23 \pm 0.03$ & $2.21 \pm 0.03$ & $2.30 \pm 0.05$ \\
    \small{Pleistochrome}       & $3.26 \pm 0.04$ & $2.22 \pm 0.04$ & $2.33 \pm 0.05$ \\
    \midrule
    \small{\textbf{Color App. Models}} & \small{CIE Lab} & \small{L Lab} & \small{CIECAM} \\
    \midrule
    \small{LMS nonlin. sensors} & $5.89 \pm 0.04$ & $5.84 \pm 0.02$ & $6.55 \pm 0.03$ \\
    \small{ATD nonlin. sensors} & $\blue{\mathbf{1.04}} \pm 0.06$ & $2.00 \pm 0.02$ & $1.58 \pm 0.05$ \\
    \midrule
    \small{\textbf{Statist. Model}} & \small{No adaptation} & \small{Von-Kries} & \small{Webster-Clifford} \\
    \midrule
    \small{Infomax SPCA}             & $1.12 \pm 0.04$ & $\blue{\mathbf{0.95}} \pm 0.06$ & $\blue{\mathbf{1.08}} \pm 0.04$ \\
  \bottomrule
\end{tabular}
\end{table}


\begin{table}
  \caption{Differential entropy of color through the networks (in bits). Upper bound for $H$ in the common cube is 9.97 bits.}
  \vspace{-0.3cm}
  \label{table_H}
  \begin{tabular}{lccc}
    \toprule
    \small{\textbf{Physiol. Models}} & \small{No adaptation} & \small{Von-Kries} & \small{Webster-Clifford} \\
    \midrule
    \small{Linear LMS input}    & $-4.97 \pm 0.04$ & $-4.01 \pm 0.04$ & $-4.01 \pm 0.02$ \\
    \small{Linear ATD channels} & $-1.68 \pm 0.03$ & $-0.71 \pm 0.03$ & $-0.72 \pm 0.01$ \\
    \small{Pleistochrome}       & $6.27 \pm 0.04$ & $7.64 \pm 0.03$ & $7.56 \pm 0.02$ \\
    \midrule
    \small{\textbf{Color App. Models}} & \small{CIE Lab} & \small{L Lab} & \small{CIECAM} \\
    \midrule
    \small{LMS nonlin. sensors} & $1.93 \pm 0.04$ & $-2.18 \pm 0.02$ & $-1.70 \pm 0.03$ \\
    \small{ATD nonlin. sensors} & $4.61 \pm 0.02$ & $1.10 \pm 0.03$ & $2.74 \pm 0.01$ \\
    \midrule
    \textbf{Statist. Model} & \small{No adaptation} & \small{Von-Kries} & \small{Webster-Clifford} \\
    \midrule
    \small{Infomax SPCA}            & $\blue{\mathbf{8.70}} \pm 0.04$ & $\blue{\mathbf{8.82}} \pm 0.03$ & $\blue{\mathbf{8.71}} \pm 0.03$ \\
  \bottomrule
\end{tabular}
\end{table}


\begin{table}
  \caption{Transferred information (input-output mutual information, in bits). Upper bound for $I$ is 14.1 bits.}
  \vspace{-0.3cm}
  \label{table_I}
  \begin{tabular}{lccc}
    \toprule
    \small{\textbf{Physiol. Models}} & \small{No adaptation} & \small{Von-Kries} & \small{Webster-Clifford} \\
    \midrule
    \small{Linear LMS input}    & $5.0 \pm 0.1$ & $5.1 \pm 0.1$ & $5.2 \pm 0.1$ \\
    \small{Linear ATD channels} & $7.7 \pm 0.2$ & $7.6 \pm 0.2$ & $7.6 \pm 0.1$ \\
    \small{Pleistochrome}       & $8.58 \pm 0.05$ & $8.6 \pm 0.1$ & $8.7 \pm 0.1$ \\
    \midrule
    \small{\textbf{Color App. Models}} & \small{CIE Lab} & \small{L Lab} & \small{CIECAM} \\
    \midrule
    \small{LMS nonlin. sensors} & $6.6 \pm 0.1$ & $5.44 \pm 0.03$ & $5.8 \pm 0.1$ \\
    \small{ATD nonlin. sensors} & $\blue{\mathbf{9.8}} \pm 0.2$ & $7.8 \pm 0.2$ & $\blue{\mathbf{8.8}} \pm 0.3$ \\
    \midrule
    \small{\textbf{Statist. Model}} & \small{No adaptation} & \small{Von-Kries} & \small{Webster-Clifford} \\
    \midrule
    \small{Infomax SPCA}                & $\blue{\mathbf{8.9}} \pm 0.2$ & $8.7 \pm 0.3$ & $\blue{\mathbf{8.8}} \pm 0.3$ \\
  \bottomrule
\end{tabular}
\end{table}


Regarding indirect measures of efficiency, representations that minimize the redundancy, $T$, and maximize the entropy, $H$, are better for signal representation for the following reasons \cite{Bell95}: independent components lead to factorial codes that maximize the use of the
channel capacity, and maximum entropy representations imply that the signal accepts more noise without significant information loss.

Good representations in terms of the indirect measures should also lead to good performance in terms of the amount of information
from the input that is available at the internal representation (the shared information, $I$).
By definition, noiseless and invertible representations preserve all the information from the input.
Therefore, the measurement of the transmitted information only makes sense for noisy sensors (which is the case of actual physiological mechanisms).
According to this, just for illustrative purposes, in our experiments measuring $I$ every considered sensor was subject to Gaussian noise whose standard deviation was 5\% of the total deviation of the response.
We computed $I$ between these noisy representations and the input (assuming a negligible noise in the input, with standard deviation
of 0.05\% of the total deviation).
As convenient reference, $I$ between the noiseless input and the negligible-noise input (i.e. the maximum available information) is $14.1 \pm 0.1$ bits. The values in Table \ref{table_I} should be compared to that upper bound.

Finally, in Table \ref{table_KLD} we report the \emph{KLD} that measures the correspondence between the color sets corresponding to D65 illumination and A illumination after chromatic adaptation (or white balance). Lower divergences indicate better match between the color compensated sets.
In all tables the three best results are highlighted in blue.

\begin{table}
  \caption{Chromatic adaptation (Kullback-Leibler Divergence between the D65 and A sets, in bits). Perfect color compensation is represented by KLD = 0 bits.}
  \vspace{-0.3cm}
  \label{table_KLD}
  \begin{tabular}{lccc}
    \toprule
    \small{\textbf{Physiol. Models}} & \small{No adaptation} & \small{Von-Kries} & \small{Webster-Clifford} \\
    \midrule
    \small{Linear LMS input}    & $5.7 \pm 0.1$ & $0.84 \pm 0.05$ & $\mathbf{\blue{0.67}} \pm 0.06$ \\
    \small{Linear ATD channels} & $3.6 \pm 0.3$ & $0.83 \pm 0.05$ & $0.78 \pm 0.04$ \\
    \small{Pleistochrome}       & $3.6 \pm 0.3$ & $0.7 \pm 0.1$ & $0.82 \pm 0.07$ \\
    \midrule
    \small{\textbf{Color App. Models}} & \small{CIE Lab} & \small{L Lab} & \small{CIECAM} \\
    \midrule
    \small{LMS nonlin. sensors} & $\mathbf{\blue{0.54}} \pm 0.07$ & $0.72 \pm 0.09$ & $0.73 \pm 0.07$ \\
    \small{ATD nonlin. sensors} & $\mathbf{\blue{0.55}} \pm 0.07$ & $0.72 \pm 0.07$ & $0.72 \pm 0.02$ \\
    \midrule
    \small{\textbf{Statist. Model}} & \small{No adaptation} & \small{Von-Kries} & \small{Webster-Clifford} \\
    \midrule
    \small{Infomax SPCA}                & $2.2 \pm 0.1$ & $1.8 \pm 0.1$ & $1.6 \pm 0.1$ \\
  \bottomrule
\end{tabular}
\end{table}

\section{Discussion}

Visualization of manifold changes through the transforms of the considered models confirms their statistical interpretation outlined in the introduction: adaptation aligns the data obtained in different acquisition conditions, linear opponent channels
rotate the input LMS representation following the axes of the data, and the nonlinear nature of the opponent channels equalizes the responses.

While data in the input representation is highly correlated due to the overlap of the LMS sensitivities (strong alignment along the diagonal of the domain in the top row of Figs. \ref{fig_physiol} and \ref{fig_cams}), the redundancy between the responses clearly reduces at later stages.
This is quantitatively confirmed by the reduction of $T$ along the layers of the networks in Table \ref{table_T}.
On the other hand, visualizations suggest that data is progressively equalized along the networks and this is quantitatively confirmed by the progressive increase of the differential entropy along the columns of Table \ref{table_H}.
As a result of the progressive independence and equalization, the amount of information about the input available at the different representations increases along the columns of Table \ref{table_I}.

It is remarkable that psychophysically-tuned models like CIE-Lab and CIECAM have similar or better information transmission performance
than an unsupervised learning method, SPCA, specifically trained for information maximization.
While the emergence of perceptual nonlinearities, adaptation, and aftereffects from SPCA \cite{Laparra12,Laparra15} is a confirmation of the Efficient Coding Hypothesis in the classical \emph{from-statistics-to-perception} direction, the quantitative efficiency of CIE Color Appearance Models presented here is an alternative confirmation in the \emph{from-perception-to-statistics} direction.

Beyond this confirmation, these results allow to quantify the gain in information transference due to the specific layers of the network:
retinal adaptation, transform to opponent channels, and saturation nonlinearities
of the opponent channels. Average results of the differences, $\Delta I$,
are given in Table~\ref{table_deltaI1}. This implies that \emph{opponency} is the most relevant feature of color vision to
favour efficient information transmission, followed by the \emph{saturating response} of the opponent channels. These processes are way more
important than \emph{chromatic adaptation}. One may argue that the goal of chromatic adaptation is making scene interpretation
more robust and not merely improving the information capacity of the visual pathway.

\begin{table}[b]
  \caption{Gains in available information ($\Delta I$, in bits) due to different features of the models}
  \vspace{-0.3cm}
  \label{table_deltaI1}
  \begin{tabular}{lccc}
    \toprule
                & \small{\textbf{Retinal Adaptation}} & \small{\textbf{Opponency}} & \small{\textbf{Saturation}} \\
    \midrule
    \small{Physiol. Models}     & $0.1 \pm 0.1$ & $\mathbf{\blue{2.5}} \pm 0.2$ & $1.0 \pm 0.2$ \\
    \midrule
  \end{tabular}
  \begin{tabular}{lcc}
                 & \small{\textbf{Retinal Adaptation}} & \small{\textbf{Opponency+Saturation}} \\
    \midrule
    \small{Color App. Mod.}    & $0.9 \pm 0.9$ & $\mathbf{\blue{2.8}} \pm 0.4$  \\
    \small{Infomax SPCA}       & $0.0 \pm 0.3$ & $\mathbf{\blue{3.7}} \pm 0.2$  \\
  \bottomrule
\end{tabular}
\end{table}

Note also that \emph{KLD} results in Table \ref{table_KLD} show that SPCA (designed for information maximization)
actually gets lower adaptation performance than simple Von-Kries or Webster-Clifford procedures in the LMS space.
This together with the small impact of chromatic adaptation in information transmission suggest that adaptation should be
explained by a different principle.


The communication efficiency analysis done here following the \emph{perception-to-statistics} logic is different
from the work of Foster et al., also concerned about the use of accurate information-theoretic measures
in color vision \cite{FosterJOSA18}. Note that their work is mainly focused on determining the number of discriminable
colors/surfaces in different illumination conditions \cite{FosterVisNeuro04,FosterJOSA09,FosterJoV10}, 
which is related to the amount of color information in a scene that can be extracted from color measurements under other illumination \cite{Ivan13,FosterJOSA18}. These problems are related to entropy and mutual-information measures, but they do not quantify the information transference through the visual pathway (mutual information between layers and redundancy within layers).
As an example, in \cite{FosterJOSA09,FosterCIC08} the redundancy is considered only because of its impact on the available information
in the color compensation context, not as a measure of information transmission of the visual system.


\section{Conclusions}

Results show that perceptual architectures reduce about 75\% of the redundancy present at the input linear responses.
As a result, while noisy sensors at the input representation would retain only 35\% of the chromatic information,
sensors with the same amount of noise at the inner representations would retain about 65\% of the information.
In terms of communication efficiency, the most relevant transform is the consideration of opponent channels
followed by the nonlinear response of the opponent channels. On the contrary, the impact of adaptation to improve the
transmission capacity is almost negligible.

From the theoretical neuroscience perspective, these results confirm the \emph{Efficient Coding Hypothesis} for
human color vision in the \emph{perception-to-statistics} direction:
statistically-agnostic color vision models such CIE Lab and CIECAM02 (only based on psychophysics)
are remarkably efficient in transmitting natural colors.
Moreover, for the data science community, these results rank the relevance of color vision features
in terms of their impact in optimal color information transmission.

\section{Acknowledgements}

This work was partially funded by the Spanish Ministerio de Economia y Competitividad
(MINECO/FEDER,UE) project DPI2017-89867-C2-2-R and by the Generalitat Valenciana grant GrisoliaP/2019/035.

\balance

\bibliographystyle{ACM-Reference-Format}

%
%
%
%
%
%
%
%
%

\end{document}